\documentclass[preprint,tightenlines,showpacs,showkeys,floatfix,nofootinbib,superscriptaddress,fleqn]{revtex4}
\usepackage{bm}
\usepackage{dcolumn}
\usepackage{amsmath}
\usepackage{amsfonts}
\usepackage{amssymb}
\usepackage{amscd}
\usepackage{amsthm}
\usepackage{graphicx}
\newcommand{\Det}{{\rm Det}}
\newcommand{\Tr}{{\rm Tr}}
\newcommand{\tr}{{\rm tr}}

\newcommand{\be}{\begin{equation}}
\newcommand{\ee}{\end{equation}}
\newcommand{\bea}{\begin{eqnarray}}
\newcommand{\eea}{\end{eqnarray}}
\newcommand{\ba}{\begin{array}{l}}
\newcommand{\ea}{\end{array}}

\newcommand{\re}[1]{(\ref{#1})}
\begin{document}
\preprint{PNU-NTG-06/2004}
\title {Magnetic susceptibility of the QCD vacuum}
\author{Hyun-Chul Kim}
\email{hchkim@pusan.ac.kr}
\affiliation{Department of Physics and  Nuclear Physics \& Radiation
Technology Institute (NuRI), Pusan National University, 609-735
Busan, Republic of Korea}
\author{M. Musakhanov}
\email{musakhanov@pusan.ac.kr} \affiliation{Department of Physics
and  Nuclear Physics \& Radiation Technology Institute (NuRI),
Pusan National University, 609-735 Busan, Republic of Korea}
\affiliation{Theoretical Physics Department, National University
of Uzbekistan, Tashkent 700174,
  Uzbekistan}
\author{M. Siddikov}
\email{Marat.Siddikov@tp2.ruhr-uni-bochum.de}
\affiliation{Theoretical Physics Department, National University
of Uzbekistan, Tashkent 700174,
  Uzbekistan}
\affiliation{Institut f\"ur Theoretische  Physik  II,
Ruhr-Universit\" at Bochum,   D--44780 Bochum, Germany}

\date{November 2004}

\begin{abstract}
We investigate the magnetic susceptibility of the QCD vacuum,
based on the instanton vacuum.  Starting from the instanton liquid
model for the instanton vacuum, we derive the light-quark partition
function $Z[V,T,\hat{m}]$ in the presence of the current quark mass
$\hat{m}$ as well as the external Abelian vector and tensor fields.
We calculate a two-point correlation function relevant for the
magnetic susceptibility and derive it beyond the
chiral limit.  We obtain for the different flavors the following
magnetic susceptibility: $\chi_{u,d}\langle i\psi_{u,d}^\dagger
\psi_{u,d}\rangle_0 \sim 40\sim45 \,{\rm MeV}$, while $\chi_{s}\langle
i\psi_s^\dagger \psi_s \rangle_0 \simeq 6\sim 10 \,{\rm MeV}$ with the 
quark condensate $\langle i\psi^\dagger \psi\rangle_0$.
\end{abstract}
\maketitle
{\bf 1.} The QCD vacuum is one of the most complicated states filled
not only by perturbative but also strong nonperturbative fluctuations.
In particular, the condensates of the quarks and gluons characterize
its nonperturbative aspect.  The quark condensate plays a role of an
order parameter associated with spontaneous chiral symmetry breaking 
(S$\chi$SB) which is probably one of the most important quantities for
low-energy hadronic phenomena.

The presence of a constant external electromagnetic field
$F_{\mu\nu}$ induces another type of condensates leading to the
nonzero magnetic susceptibility $\chi_f$ of the QCD vacuum defined as:
\begin{equation}
\langle0|\psi_f^\dagger \sigma_{\mu\nu}\psi_f|0\rangle_F = e_f
\,\chi_f\, \langle i\psi_f^\dagger \psi_f \rangle_0\,F_{\mu\nu}, 
\label{chi}
\end{equation}
where $e_f$ denotes the quark electric charge, $f$ is the
corresponding flavor.  It is natural to have the quark
condensate $\langle i\psi_f^\dagger \psi_f \rangle_0 $ in the chiral
limit as a normalization factor in the right-hand side of Eq.
\re{chi}, since S$\chi$SB is responsible for these quantities,
i.e. the quark condensate and magnetic susceptibility.  In a recent
paper~\cite{Braun:2002en} it was suggested that the magnetic
susceptibility $\chi_f$ may be measured in the exclusive
photoproduction of hard dijets $\gamma+N\rightarrow (\bar qq)+N$.  The
value of $\chi_f$ was predicted by using the method of the QCD sum
rule and vector dominance~\cite{Belyaev:ic,Balitsky:aq,Ball:2002ps}:
$\chi_f \langle i \psi_f^\dagger \psi_f \rangle_0 = 40 \sim 70\, {\rm
MeV}$ at the scale of $1\, {\rm GeV}$.  

In the present work, we want to investigate the magnetic
susceptibility $\chi_f$ of the QCD vacuum within the framework of the 
instanton vacuum.    Since the instanton vacuum explains S$\chi$SB
naturally via quark zero modes, it may provide a good framework to
study the $\chi_f$.  Moreover, there are only two parameters in this
approach, namely the average instanton size $\rho\approx \frac{1}{3}\,
{\rm fm}$ and average inter-instanton distance $R\approx 1\, {\rm
  fm}$.  The normalization scale of this approach can be defined by
the average size of instantons and is approximately equal to
$\rho^{-1}\approx 0.6\, {\rm GeV}$.  The values of the $\rho$ and $R$
were estimated many years ago phenomenologically by
Ref.~\cite{Shuryak:1981ff} as well as theoretically by
Ref.~\cite{Diakonov:1983hh}.  Furthermore, it was recently confirmed in
various lattice simulations of the QCD vacuum
\cite{Chu:vi,Negele:1998ev,DeGrand:2001tm}.  Very recent lattice
calculations of the quark
propagator~\cite{Faccioli:2003qz,Bowman:2004xi} are in a remarkable  
agreement with that of Ref.~\cite{Diakonov:1983hh}. 

The magnetic susceptibility of the QCD vacuum was estimated already in
Ref.~\cite{Petrov:1998kg}, however, done in the chiral limit and
without the consideration of the current conservation.  Since we want
to investigate it beyond the chiral limit, we first need to extend  
Refs.~\cite{Diakonov:1985eg,Diakonov:1995qy,Diakonov:2002fq} and 
Lee-Bardeen formula~\cite{Lee:sm} for a low-frequency ($p \ll
\rho^{-1}$) part of the quark determinant:
 \begin{equation}
{\Det}_{\rm low}=\det B, \,\, B_{ij}= i\hat{m}\delta_{ij} + a_{ji},
\,\,\, a_{-+}=-\langle \Phi_{- , 0} | i\rlap{/}{\partial} |\Phi_{+, 0}
\rangle, 
\label{detB}
\end{equation}
taking into account the current quark mass.  Here, $\hat{m}$ denotes a
current quark mass, which is assumed to be small, $a_{ij}$ 
stands for the overlapping matrix element of the quark zero modes
$\Phi_{\pm, 0}$ generated by instantons (antiinstantons).  In
general, we define $\Phi_{\pm, n}$ by the following equation:
\begin{equation}
(i\rlap{/}{\partial} + \rlap{/}{A}_{\pm})
\Phi_{\pm,n}=\lambda_{n}\Phi_{\pm,n}, 
\end{equation}
where $A_{\mu, \pm}$ is an instanton (antiinstanton) field and
$\lambda_0 = 0$ by convention.  The matrix element $a_{ij}$ is
nonzero only between instantons and antiinstantons and vice versa
due to specific chiral properties of the zero modes.

The overlapping of the quark zero modes allows quarks to propagate
through the instanton-antiinstanton medium by jumping from one
instanton to another one.  Hence, ${\Det}_{\rm low}$ is reduced to the
determinant of the finite matrix $B$ in the subspace of {\em
zero modes}.  This result was reproduced by
Refs.~\cite{Diakonov:1995qy,Schafer:1996wv} in different ways.
It is well known that the fermionic determinant is proportional to
$m^{|N_{+}-N_{-}|}$ and strongly suppresses the fluctuations of
$|N_{+}-N_{-}|$.  Therefore, we can take $N_{+}=N_{-}=N/2$ in the
final formulas.

In order to calculate the magnetic susceptibility, we have to find
the light-quark partition function $Z[V,T,\hat{m}]$ in the presence of
the Abelian external vector $V_\mu$ and antisymmetric tensor
$T_{\mu\nu}$ fields.  So, we first calculate a zero-mode approximated
quark propagator $\tilde S$ in the instanton ensemble and in
the presence of these external fields.  Using this propagator, we are
able to compute a low-frequency part of the quark determinant 
$\tilde{\Det}_{\rm low}$ and arrive at the extension of Eq. \re{detB}
for the non-zero $\hat m$ with the external Abelian vector $V$ and
tensor $T$ fields.

The smallness of the packing parameter $\pi(\frac{\rho}{R})^4\approx
0.1$ makes it possible to average the determinant over
collective coordinates of instantons with fermionic quasiparticles,
i.e. constituent quarks $\psi$ introduced.  The averaged 
determinant turns out to be the light-quark partition function
$Z[V,T,\hat{m}]$ which is a functional of $V$ and $T$ and can be
represented by a functional integral over the constituent quark fields
with the gauged effective chiral action 
$S[\psi^\dagger,\psi,V,T]$.  However, it is not trivial to make the
action gauge-invariant due to the nonlocality of the quark-quark
interactions generated by instantons.  In the previous
paper~\cite{Musakhanov:2002xa}, it was demonstrated how to gauge the
nonlocal effective chiral action in the presence of the external
electromagnetic field.  The gauged action was successfully tested by
showing that the low-energy theorem of the axial anomaly related to
the process $G\tilde G\rightarrow \gamma\gamma$ (see also
\cite{Musakhanov:1996qf,Salvo:1997nf}) is satisfied.  

In the present work we refine the derivation of the gauged effective
chiral action and apply it to the calculations of the magnetic
susceptibility $\chi_f$ of the QCD vacuum, taking into account both
current quark mass and nonlocal currents.


{\bf 2.} It is clear that the magnetic susceptibility $\chi_f$ is
related to the two-point correlation function of the quark tensor and 
vector currents at the zero momentum.  In order to calculate this 
correlation function, we need to treat the light-quark partition function
$Z[V,T,\hat{m}]$ as a functional of external abelian vector $V_\mu$
and tensor $T_{\mu\nu}$ fields.  The relevant total quark determinant 
\begin{equation}
\tilde{\Det}:= \Det(i\rlap{/}{\partial}  + \rlap{/}{A} + e\rlap{/}{V}
+ \sigma_{\mu\nu} T_{\mu\nu} + i\hat{m}) 
\end{equation}
can be splitted into two parts which correspond to low and high
momenta, respectivley, with respect to some auxiliary parameter $M_1$
lying inside an interval, $R^{-1} \ll M_1 \ll \rho^{-1}$: $\tilde
{\Det} = \tilde{\Det}_{\rm low}\times\tilde{\Det}_{\rm high}$
~\cite{Diakonov:1985eg}.  $\tilde{\Det}_{\rm high}$ comes
from fermion modes with Dirac eigenvalues from the
interval $M_1$ to the Pauli--Villars mass $M$, and
$\tilde{\Det}_{\rm low}$ accounts for eigenvalues less than
$M_1$.  The product of these determinants is independent of the scale
$M_1$.  However, we can deal with them only approximately.  The
high-momentum part $\tilde{\Det}_{\rm high}$ can 
be written as a product of the determinants in the field of individual
instantons, while the low-momentum one ${\tilde\Det}_{\rm low}$ has to
be treated approximately, would-be zero modes being only taken into 
account.  In Ref.~\cite{Diakonov:1985eg}, it was demonstrated that the
${\tilde\Det}_{\rm low}$ depends weakly on $M_1 $ in the wide range,
which serves as a check for this approximation.  Based on
this prescription, we now calculate the low-momentum part of the quark
determinant in the presence of instantons as well as external $V_\mu
$, $T_{\mu\nu}$ fields, and then average it over the collective
coordinates of the instantons in order to find the effective
low-energy QCD partition function as a functional of the external
field $Z[V,T,\hat{m}]$.

We first define the total quark propagator $\tilde S $ in the
presence of the instanton ensemble $A$ and external fields $V$,
$T$ and the quark propagator $\tilde S_i$ with a
single instanton $A_i$ as well as $V$ and $T$:
\begin{equation}
\tilde S = \frac{1}{i\rlap{/}{\partial} + g \rlap{/}{A}  +e\rlap{/}{V}
+\sigma\cdot T +i\hat{m}},\;\;\;
\tilde S_i=  \frac{1}{i\rlap{/}{\partial} + g \rlap{/}{A_i}  +e\rlap{/}{V}
+\sigma\cdot T +i\hat{m}},
\end{equation}
where $\sigma\cdot T=\sigma_{\mu\nu}T_{\mu\nu}$.  We assume that the
total instanton field $A$ may be approximated as a sum of the single
instanton fields, $A=\sum_{i=1}^{N}A_i$, which is justified with the 
above-mentioned average size of instantons $\rho\approx 1/3 \,{\rm
  fm}$ and average inter-instanton distance $R \approx 1 \, {\rm fm}$.
Defining the quark propagator $\tilde S_0$ with external fields $V$
and $T$ and the free one $S_0$ as follows:
\begin{equation}
\tilde S_0=\frac{1}{i\rlap{/}{\partial} + e\rlap{/}{V}+\sigma\cdot
  T +i\hat{m}} ,\,\,\,
S_0=\frac{1}{i\rlap{/}{\partial} +i\hat{m}},
\end{equation}
we can expand the quark propagator $\tilde S$ with respect to a single
instanton:
\bea
\tilde S=\tilde S_0+\sum_i (\tilde S_i-\tilde
S_0)+\sum_{i\not=j} (\tilde S_i-\tilde S_0)\tilde S^{-1}_0(\tilde
S_j-\tilde S_0)+\cdots .
\label{S-tot}
\eea
In order to specify the gauge dependence, we rewrite $\tilde S_i$ and
$S_0$ in the following form:
\begin{eqnarray}
\tilde S_i&=&L_iS'_{i}L^{-1}_i,\;\; S'_{i}=\frac{1}{i\rlap{/}{\partial}
+g\rlap{/}{A_i} +e\rlap{/}{V'}+\sigma\cdot T
+i\hat{m}},\cr
\tilde S_0&=&L_iS'_{0}L^{-1}_i,\;\;
S'_{0i}=\frac{1}{i\rlap{/}{\partial} +  e\rlap{/}{V'}+\sigma\cdot T +
  i\hat{m}}, 
\end{eqnarray}
where $e\rlap{/}{V^\prime}=L^{-1}(i\rlap{/}{\partial}+e\rlap{/}{V})L$ and the
gauge connection $L$ can be written as the path-ordered exponent:
\begin{equation}
L_i(x,z_i)={\rm P} \exp\left(ie\int_{z_i}^x d\xi_\mu V_\mu(\xi)\right),
\label{transporter}
\end{equation}
where $z_i$ denotes an instanton position.

In the case of the small quark mass $\hat{m}$, we can expand the quark
propagator $S_i$ ~\cite{Diakonov:1985eg} as follows:
\begin{equation}
S_i(\hat{m}\rightarrow 0) =\frac{1}{i\rlap{/}{\partial}} +
\frac{|\Phi_{0i}\rangle\langle\Phi_{0i}|}{i\hat m}.
\end{equation}
While it gives a proper value for the $ \langle\Phi_{0i}|S_i
(\hat{m}\rightarrow 0)|\Phi_{0i}\rangle =\frac{1}{i\hat m}$, the second
term has wrong chiral properties:
\begin{equation}
S_i (\hat{m}\rightarrow
0)|\Phi_{0i}\rangle = \frac{|\Phi_{0i}\rangle}{i\hat m}
+\frac{1}{i\rlap{/}{\partial}}|\Phi_{0i}\rangle.
\end{equation}
However, we may neglect this term as $\hat{m}\rightarrow 0.$

If the $\hat{m}$ is not small, then we propose the following approximation:
\begin{equation}
S_i=S_{0} + S_{0}i\rlap{/}{\partial}
\frac{|\Phi_{0i}\rangle\langle\Phi_{0i}|}{c_i} i\rlap{/}{\partial} S_{0},
\label{proposed}
\end{equation}
where
\begin{equation}
c_i=-\langle\Phi_{0i}|i\rlap{/}{\partial} S_{0} i\rlap{/}{\partial}
|\Phi_{0i}\rangle = i\hat{m} \langle\Phi_{0i}|S_{0}i\rlap{/}{\partial} 
|\Phi_{0i}\rangle.
\end{equation}
The merit of the approximation proposed in Eq.(\ref{proposed}) lies 
in projecting $S_i$ onto the zero-mode:
\begin{equation}
S_i|\Phi_{0i}\rangle = \frac{1}{i\hat{m}}|\Phi_{0i}\rangle,\,\,\,
\langle\Phi_{0i}|S_i =\langle\Phi_{0i}|\frac{1}{i\hat{m}}.
\end{equation}

Including the external fields $V$ and $T$, we get
\begin{equation}
\tilde S_i =\tilde S_{0i} + \tilde S_{0i}L_ii\rlap{/}{\partial}
\frac{|\Phi_{0i}\rangle\langle\Phi_{0i}|}{c_i-b_i} i\rlap{/}{\partial}
L^{-1}_i\tilde S_{0i},
\end{equation}
where
\begin{equation}
b_i -c_i=\langle\Phi_{0i}|i\rlap{/}{\partial} (L_i^{-1}\tilde S_{0}L_i
) i\rlap{/}{\partial} |\Phi_{0i}\rangle.
\end{equation}
Thus, we can express Eq.\re{S-tot} as
\be
\tilde S =\tilde S_{0} + \tilde S_{0}\sum_{i,j}L_ii\rlap{/}{\partial}
|\Phi_{i0}\rangle
\left(\frac{1}{-D-T}\right)_{ij}\langle\Phi_{0j}|i\rlap{/}{\partial}
L_j^{-1}\tilde S_{0},
\label{propagator}
\ee
where
\begin{equation}
D_{ij}+T_{ij}=\langle\Phi_{0i}|i\rlap{/}{\partial} (L_i^{-1}\tilde
S_{0}L_j )i\rlap{/}{\partial} |\Phi_{0j}\rangle,\,\,\, D_{ij}=
(b_i-c_i)\delta_{ij}.
\end{equation}
Introducing the following function:
\begin{equation}
|\phi_{0i}\rangle=\frac{1}{i\rlap{/}{\partial}}L_i i\rlap{/}{\partial}
|\Phi_{0i}\rangle,
\end{equation}
which  has the same chiral properties as the zero-mode eigenfunction 
$|\Phi_{0i}\rangle$, and taking the trace over 
the subspace of the zero-modes, we obtain the following expression: 
\begin{equation}
\tilde\Tr (\tilde S -\tilde S_0 )= -\sum_{i,j}
\langle\phi_{0,j}|i\rlap{/}{\partial}\, ({\tilde S_{0}}^2)\,
i\rlap{/}{\partial} |\phi_{0,i}\rangle \langle\phi_{0,i}|(\frac{1}
{i\rlap{/}{\partial} \tilde S_{0}i\rlap{/}{\partial}}) |\phi_{0,j}\rangle,
\end{equation}
where $\tilde \Tr$ denotes the trace over the zero-mode subspace
only.  Bringing now in the matrix
\bea
\tilde B(m)_{ij}= \langle\phi_{0,i}|(i\rlap{/}{\partial} \tilde S_0
 i\rlap{/}{\partial})
|\phi_{0,j}\rangle ,
\label{tildeB0}
\eea
we get
\be
\tilde\Tr\int^m_{M_1}idm'(\tilde S(m')- \tilde S_{0}(m'))
 = \tilde\Tr \int^{\tilde B(m)}_{\tilde B (M_1)}
d \tilde B(m')\frac{1}{\tilde B(m')} =\tilde\Tr \ln \frac{\tilde
B(m)}{\tilde B(M_1)},
\ee
from which we find that $\tilde B=\tilde B(m)$ is equivalent to a mere 
extension of the Lee-Bardeen matrix $B$ ~\cite{Lee:sm} in the
presence of the external Abelian fields $V$, $T$ and current quark
mass $\hat{m}$.  Thus, we derive for arbitrary flavor $N_f$
\begin{equation}
  \label{eq:tildeB1}
\tilde B^{f}_{ij}= \langle\Phi_{0i}|i\rlap{/}{\partial} L_{fi}^{\dagger} \,
\tilde S_{0f}L_{fj} i\rlap{/}{\partial} |\Phi_{0j}\rangle 
\end{equation}
\begin{equation}
\tilde S_{0f}=\frac{1}{i\rlap{/}{\partial} + e_f\rlap{/}{V} + \sigma
  \cdot T +i m_f},\,\,\, L_{fi}= {\rm P} \exp\left(ie_f\int_{z_i}^x
  d \xi_\mu V_\mu(\xi)\right).
\end{equation}
If we introduce Grassmanian variables $\bar\Omega_{i}^{f}$ and
$\Omega_{j}^{g}$, then we can express the fermionic determinant of $B$
as a functional integral over fermion fields:
\begin{eqnarray}
\label{omegatildeB}
\det\tilde  B &=& \int d\Omega d\bar\Omega \exp\left[ (\bar\Omega\tilde B
\Omega ),\, (\bar\Omega \tilde B \Omega )\right]  =
\bar\Omega_{i}^{f}  \langle\phi_{0,i}|i\rlap{/}{\partial} \tilde S_{0, f}
i\rlap{/}{\partial}|\phi_{0,j}\rangle \Omega_{j}^{f} \cr
 &=&\left(\Det(i\rlap{/}{\partial} +e\rlap{/}{V}
  +\sigma\cdot T +i\hat{m})\right)^{-1} 
\int \prod_{i, f} d\Omega_{i}^{f} d\bar\Omega_{i}^{f} D\psi_f
D\psi^{\dagger}_{f} \cr
&\times& {\rm e}^{\int dx \left\{\psi^{\dagger}_{f} (x) 
(i\rlap{/}{\partial} + e_f \rlap{/}{V}+ \sigma \cdot T \,+\,i
m_f)\psi_{f} (x)+ \bar\eta_{i}^{f} (x)\psi_{f}
(x) \,+\,\psi^{\dagger}_{f} (x)\eta_{i}^{f} (x)\right\} },
\end{eqnarray}
where the source fields $\eta_{i}^{f}$ and
$\bar\eta_{j}^{f}$ are defined as:
\begin{equation}
\bar\eta_{i}^{f}= \bar\Omega_{i}^{f}
\langle \phi_{0,i}|i\rlap{/}{\partial} ,\;\; \eta_{j}^{f} =
i \rlap{/}{\partial}|\phi_{0,j} \rangle \Omega_{j}^{f}.
\label{eta}
\end{equation}
The integration over the Grassmanian variables $\Omega$ and
$\bar\Omega$ provides finally the low-frequency part of the quark
determinant in the following form:
\bea
\label{part-func}
 &&\tilde{\Det}_{\rm low} =\det\tilde B =
 \left(\det(i\rlap{/}{\partial} +e\rlap{/}{V} +\sigma \cdot T +
 i\hat{m})\right)^{-1} \int \prod_{f}D\psi_f D\psi^{\dagger}_{f}
\\\nonumber
&\times&
 {\rm e}^{\left(\int d^4 x
\psi_{f}^{\dagger} (i\rlap{/}{\partial} \,+\,e_f\rlap{/}{V}
 \,+\sigma \cdot T\, +\, im_f )\psi_{f}\right)}
 \prod_{f}\left\{\prod_{+}^{N_{+}}
V_{+,f}[\psi_{f}^{\dagger},\psi_f ]
\prod_{-}^{N_{-}}V_{-,f}[\psi_{f}^{\dagger},\psi_f ]\right\}\; ,
\eea
where
\bea
\tilde V_{\pm,f}[\psi_{f}^{\dagger},\psi_f ]=\int d^4 x
\left(\psi_{f}^{\dagger} (x)\,L_{f}(x,z) i\rlap{/}{\partial}
\Phi_{\pm, 0} (x; \xi_{\pm})\right) \int d^4 y \left(\Phi_{\pm ,
0} ^\dagger (y; \xi_{\pm} ) (i\rlap{/}{\partial} L^{+}_{f}(y,z)
\psi_{f} (y)\right).
\label{tildeV}
\eea
It is obvious that $\tilde{V}_{\pm, f}[\psi^{\dagger} ,\psi ]$
indicates the nonlocal interaction between constituent quarks
generated by instantons.  Since the range of the integration in
Eq.(\ref{tildeV}) is truncated at $\rho$, which is defined
by zero-mode functions $ \Phi_{\pm, 0}$, the range of the nonlocality  
is determined by $\rho$.

Note that the external vector field $V_\mu$ gauges not only the
kinetic term of the effective action but also its interaction one
$\tilde V_{\pm, f}[\psi_{f}^{\dagger} ,\psi_f ]$ in
Eq.~(\ref{tildeV}).  The reason is clear: Since the instanton-induced 
interaction is nonlocal, the gauge connection $L_f$ must be attached
to each fermionic line so that the effective action may be
gauge-invariant.  Unfortunately, we encounter an ambiguity arising
from this gauge connection, since it is path-dependent in general.
However, we will show now that the most preferable path can be found
in the present case.


{\bf 3.} In order to solve the problem of gauge dependence due to the
nonlocal interaction induced by instantons, we have employed the gauge 
connection $L_f$.  Though it causes in general some arbitrariness
due to its path dependence, we can prove that such dependence can be 
minimized in the present work.  To be more specific, we consider the
extended zero mode:
\bea
(i\rlap{/}{D}+\rlap{/}{V})| \tilde\Phi_0\rangle = 0,\,\,\,
i \rlap{/}{D} = i\rlap{/}{\partial} + \rlap{/}{A}, \,\,\, |
\tilde\Phi^{(1)}_0\rangle = |\Phi_0\rangle - S_{NZ}\rlap{/}{V} |
\Phi_0\rangle,
\label{extended}
\eea
where $A$ denotes an instanton field located at $z$,
$|\tilde\Phi^{(1)}_0\rangle$ stands for the solution to order ${\cal
  O}(V)$, and $S_{NZ}$ the well-known non-zero mode of
the propagator in the instanton field (see the review
\cite{Shuryak:1981ff} and references therein).  Putting the
gauge connection $L(x,z)$ defined in Eq.~\re{transporter} into
Eq.~\re{extended}, we obtain:
\begin{equation}
(i \rlap{/}{D}+\rlap{/}{V}')|\Phi_0'\rangle = 0, \,\,\,
V'_{\mu}=L^{-1}(i\partial_{\mu}+V_{\mu})L , \,\,\,|\Phi_0'\rangle
= L^{-1} (x,z) |\tilde\Phi_0\rangle.
\label{eq:path1}
\end{equation}
$i\rlap{/}{D}\,|\Phi_0\rangle=0$ and
$S_{NZ}\,i\rlap{/}{D}=1-|\Phi_0\rangle\langle\Phi_0|$ being
used, the solution $|\Phi^{(1)'}_0\rangle$ to order ${\cal O}(V)$ in
Eq.~(\ref{eq:path1}) can be reduced to the corresponding
solution $|\tilde\Phi^{(1)}_0\rangle$ to order ${\cal O}(V)$ without any
trace of the path dependence arising from the gauge connection $L$.

However, if the zero-mode approximation $S_{NZ}\approx
\frac{1}{i\rlap{/}{\partial}}=S_{00}$ is used to find the solution to
order ${\cal O}(V)$
\begin{equation}
|\Phi^{(1)'}_{00}\rangle = |\Phi_0\rangle - S_{00}\rlap{/}{V}'|
\Phi_0\rangle,
\label{eq:phi00}
\end{equation}
then applying the inverse gauge connection $L^{-1}$ to
Eq.(\ref{eq:phi00}) leads to 
\begin{equation}
|\tilde\Phi^{(1)}_{00}\rangle= |\Phi_0\rangle -S_{00} \rlap{/}{V} |
\Phi_0\rangle + S_{00} (i\int d\xi_{\mu}V_{\mu}(\xi)) | \Phi_0\rangle,
\label{12}
\end{equation}
which depends on the path we choose explicitly.  Thus, it is the best
way that we choose the path minimizing
$\| \tilde\Phi^{(1)}_0-\tilde\Phi^{(1)}_{00}\|^2$.   Expressing
the external vector field in a Fourier form:
\begin{equation}
V_{\mu}(x)=  \int d^4 q V_{\mu}(q) e^{iq\cdot x},
\end{equation}
we can parameterize the path from $x$ to $z$ as follows:
\begin{equation}
\xi_{\mu}(x,z,t)=z_{\mu}+(x-z)_{\mu}t+p_{\mu}(x,z,t)
\end{equation}
with parameter $t\in(0,1)$ and boundary conditions
$p_{\mu}(t=0)=p_{\mu}(t=1)=0$.  It is the most
optimized path we can take, since $p_{\mu}$ should depend only on the
path difference $x-z$ in order to satisfy the conservation of the
linear momenta.  More explicitly, we write down the path in the
following form:
\begin{equation}
\xi_{\mu}(x,z,t)=z_{\mu}+(x-z)_{\mu}t+p_{\mu}t(1-t) .
\end{equation}
Then we have
\begin{equation}
i\int d\xi_{\mu}V_{\mu}(\xi)=i \int d^4 q V_\mu(q)e^{iq\cdot
  z}\int^1_0 dt[(x-z)_{\mu}+(1-2t)p_\mu]e^{iq\cdot[(x-z)t+ipt(1-t)]}.  
\label{eq:path2}
\end{equation}
In principle, we can express Eq.~(\ref{eq:path2}) in terms of the
error function.  However, we shall consider a simple case of
$q\rho\ll 1$ so that we can expand the vector current with respect to
$(q\cdot p)$ and $q\cdot (x-z)$:
\begin{equation}
V_{\mu}(x)= \int d^4 q
V_{\mu}(q)e^{iq\cdot x}\approx \int d^4 q V_\mu(q)e^{iq\cdot
z}\left[1+iq\cdot (x-z)-\frac{(q\cdot (x-z))^2}{2}\right]
\end{equation}
Substituting this expansion into Eqs.~\re{extended} and \re{12} and
carrying out some tedious but straightforward calculations, we obtain
\begin{equation}
\| \tilde\Phi^{(1)}_0- \tilde\Phi^{(1)}_{00}\|^2\approx C
\rho^2 N_f \int d^4 q V^2(q)(q \cdot p)^2+\cdots
\end{equation}
where the constant $C=+0.0063 >0$ and higher-order terms are
neglected.  Therefore, we can conclude that the minimum is achieved by
taking the straight-line path that will be employed in the sequel.

Moreover, since we are only interested in the case of a constant
electromagnetic field strength $F_{\mu\nu}$, we can even show that the
magnetic susceptibility does not depend at all on the path.  Let us
consider the following difference: 
\begin{equation}
\delta \left(\int_{z}^x d\xi_\mu V_\mu(\xi)\right) := \int_{C}
d \xi_{\mu} V_\mu(\xi) - \int_{s-l} d\xi_{\mu} V_\mu(\xi),
\end{equation}
where $\int_{C} d\xi_{\mu} V_\mu(\xi)$ is the line integral
along the path $C$ parameterized by
$\xi_{\mu}(x,z,t)=z_{\mu}+(x-z)_{\mu}t+a_{\mu}(x-z,t) $ and
$\int_{s-l} d\xi_{\mu} V_\mu(\xi)$
is calculated along the straight-line.  This difference is reducible
to the integral over the surface between these two contours due to
Stoke's theorem:
\begin{equation}
\delta \left(\int_{z}^x d\xi_\mu V_\mu(\xi)\right) := \oint d\xi_{\mu} 
V_\mu(\xi) = \oint dS_{\mu\nu} F_{\mu\nu}(\xi) = F_{\mu\nu} \int
d\xi_{\mu} a_{\nu}(\xi), 
\end{equation}
where the integral $\int d\xi_{\mu} a_{\nu}(\xi) $ depends only on a 
distance $(x-z)$ but not on $x$ and $z$ separately due to
translational invariance.  Denoting $G_{\mu\nu}(x-z)\equiv \int
d\xi_{\mu} a_{\nu}(\xi)F(x-z)$, where $F(x-z)$ represents the
form factor $F$ in coordinate space, we find that the change of the
magnetic susceptibility $\delta \chi_f$ due to the deformation
of the contour turns out to vanish:
\begin{equation}
F_{\lambda\rho} \Tr[S(p)G_{\rho\lambda}(p)F(p)
S(p)\sigma_{\mu\nu}+S(p)F(p)G_{\rho\lambda}(p)S(p)\sigma_{\mu\nu}] =0 .
\end{equation}
Thus, we are now allowed to take any path without changing the final
results of the magnetic susceptibility.


{\bf 4.} We are now in a position to derive the relevant partition
function for the magnetic susceptibility of the QCD vacuum.  We
first have to average the low-frequency part of the quark
determinant $\tilde{\det}_{\rm low}$ over collective coordinates
$\xi_\pm$. Having exponentiated and bosonized the interaction terms 
\cite{Diakonov:1985eg,Diakonov:1995qy,Musakhanov:2002xa,Musakhanov:1998wp,
Musakhanov:2001pc}, we arrive at the effective chiral action with
external fields pertinent to the magnetic susceptibility.  Since
the density of the instanton medium
($\pi^2\left(\frac{\rho}{R}\right)^4\approx 0.1$) is low, we are
able to average over positions and orientations of the
instantons independently.  In the next step we have to introduce
auxiliary fields, i.e. the couplings $\lambda_\pm$ for
exponentiation and meson fields $\Phi_\pm (x)$ for bosonization,
respectively.  Having integrated over fermion fields, we end up with
the partition function in the presence of the external vector and 
tensor fields:
\begin{equation}
Z[V,T,\hat{m}] =\int d\lambda_{+} d\lambda_{-}
 D\Phi_{+}D\Phi_{-}\exp\left(-S[\lambda_{\pm},\Phi_{\pm}]\right), 
\label{Z}
\end{equation}
where
\bea
 S[\lambda_{\pm},\Phi_{\pm}] &=& -\sum_{\pm}
 \left[ N_{\pm} \ln \left\{\left(  \frac{4\pi^2
\rho^2}{ N_c} \right)^{N_f}\frac{N_{\pm}}{V\lambda_{\pm}}\right\} -
N_{\pm}\right]+ S_{\Phi} + S_{\psi},
\label{S}
\\
 S_{\Phi} &=& \int d^4 x \sum_{\pm}
 (N_f - 1) \lambda_{\pm}^{-\frac{1}{N_f - 1}}
(\det\Phi_{\pm}(x) )^{\frac{1}{N_f - 1}}   ,
\nonumber
\\
S_{\psi} &=& -{\rm Tr} \ln ((\rlap{/}{P}  +\sigma\cdot T + i\hat{m} +
i F(P) \sum_{\pm} \Phi_{\pm }(x)\frac{1\pm\gamma_{5}}{2})F(P)
 (\rlap{/}{P} +\sigma\cdot T + i \hat{m} )^{-1}).
\nonumber
\eea
Here, $P_\mu =p_\mu +eV_\mu$, $p_\mu =i\partial_\mu$ and ${\rm
Tr}$ stands for the functional trace, i.e. $\int d^4 x
\tr_c\tr_D\tr_f$ and $F(k)$ denotes the quark form factor
generated by the fermionic zero modes and has an explicit form as
follows~\cite{Diakonov:1985eg}:
\begin{equation}
F(k\bar{\rho}) \;=\; 2z \left(I_0 (z) K_1 (z)- I_1 (z) K_0 (z)
-\frac{1}{z} I_1 (z) K_1 (z)\right),
\label{eq:ff}
\end{equation}
where $I_0$, $I_1$, $K_0$, and $K_1$ are the modified Bessel
functions, $z$ is defined as $z=k\bar{\rho}/2$.  When $k$ goes to
infinity, this zero-mode form factor $F(k\bar{\rho})$ has
the following asymptotic behavior:
\begin{equation}
F(k\bar{\rho}) \longrightarrow \frac{6}{(k\bar{\rho})^3}.
\end{equation}

In fact, the nonlocal effective quark-meson interaction can be
understood without relying on the instanton
vacuum~\cite{Ripka}.  In those cases, the momentum-dependent quark
mass can be interpreted as a nonlocal regularization in Euclidean
space.  Hence, various types of the $F(k)$ as a regulator with  the
regularization parameter $\Lambda\sim 1/ \bar{\rho}$ has been used by
different authors.  For example, the dipole-type $F(k)$ is adopted in
the study of the pion wave function~\cite{Petrov:1998kg}, while the
Gaussian is employed in Ref.~\cite{Golli:1998rf}.  Hence, we utilize
in this work the dipole and exponential types of the form factors in
addition to the instanton-induced zero-mode form factor:
\begin{eqnarray}
F(k)=\left\{\begin{array}{lll}
\Lambda^2/(\Lambda^2+k^2) & {\rm (dipole)} , \\
\exp(-k^2/\Lambda^2) & {\rm (Gaussian)}.
\end{array}\right.
\label{eq:regulator}
\end{eqnarray}

The saddle-point approximation of the integrals in the partition
function given in Eq. \re{Z} leads to the equation for the effective
quark mass, which is in general expressed by the functional of the
external fields $ V$ and $T$.  However, since we finally need the term
of order ${\cal O}(VT)$,  we can consider the saddle-point
approximation without external fields:
\begin{equation}
\Phi_{\pm ,fg}= \Phi_{\pm,fg}(0)=M_{f}\delta_{fg},
\,\,\,\,\lambda_{\pm}= \lambda = \frac{2V}{N}\prod_{f} {M_f} ,
\end{equation}
which satisfies the following saddle-point equation
\begin{equation}
\frac{N}{V} = 4N_c \int \frac{d^4 k}{(2\pi )^4}
\frac{ M_f F^{2}(k)(m_f + M_f F^{2}(k))}{k^2 + (m_f + M_f F^{2}(k) )^2}.
\label{saddle}
\end{equation}
Equation~(\ref{saddle}) contains the momentum-dependent dynamic
quark mass $M_{f}(k)=M_{f}F^{2}(k)$.  Note that the dynamic quark
mass $M_f$ is a decreasing function of the current quark mass
$m_f$ \cite{Musakhanov:1998wp,Musakhanov:2001pc} with meson loops
neglected.   At small $m_f$, we may write $M_f= M_0+\gamma m_f$
with $M_0 \approx 360$ MeV and $\gamma\approx -2$\footnote{ Our
preliminary estimate indicates that the contribution of the  meson
loops to the dynamic quark mass is not at all small due to the
chiral log corrections which are of order
$\frac{1}{N_c}\,m_f\,\ln\, m_f $.  The meson-loop corrections are 
even more important than those of order ${\cal O} (m_f)$.}.

Having derived the partition function in Eq.~(\ref{S}), we are now
ready for the calculation of the magnetic susceptibility $\chi_f$ from
the instanton vacuum within our framework.  In order to evaluate
$\chi_f$, we have to compute the following correlation function via
the functional derivative with respect to the external tensor field: 
\begin{equation}
\left.\frac{\delta}{\delta T_{\mu\nu}}\ln Z[V,T,\hat{m}] \right|_{T =
  0} = \langle 0| \psi^+ \sigma_{\mu\nu}\psi|0\rangle_F = {\rm Tr}
(\tilde S\sigma_{\mu\nu})
\end{equation}
with
\begin{equation}
\tilde S=\frac{1}{\rlap{/}{P} + i \hat{m} + i M(P)}.
\end{equation}
We shall consider here only the leading order in the large $N_c$
expansion.  Using the Schwinger
method~\cite{Schwinger:nm,Vainshtein:xd}, we solve the following
traces: 
\bea
&&\Tr\left\{\left(\frac{1}{\rlap{/}{P} + i(\hat{m}+ M(P))}
 -\frac{1}{\rlap{/}{P} + i \hat{m}}\right) \sigma_{\mu\nu}\right\} \cr
&=& i\Tr\left\{ \left(\frac{(\hat{m}+ M(p))}{(p^2 + (\hat{m}+
M(p))^2)^2}- \frac{\hat m}{(p^2 + \hat{m}^2)^2}\right)\frac{\sigma
\cdot F}{2} \sigma_{\mu\nu}\right\} \cr 
&+& i\Tr\left\{ \frac{1}{(p^2 +
(\hat{m}+ M(p))^2)^2} \rlap{/}{p}[\rlap{/}{P}
,M(P)]\sigma_{\mu\nu}\right\},
\eea
where $\sigma\cdot F=\sigma_{\mu\nu}F_{\mu\nu}$.  We need to calculate
the commutator $[\rlap{/}{P} ,M(P)]$, keeping only
terms to order ${\cal O}(V)$.  It can be easily computed for the
dipole-type form factor.  Otherwise, we can derive them
numerically.  Finally, we arrive at the following expression for the
magnetic susceptibility:
\bea
\chi_f \langle i\psi_f^\dagger \psi_f \rangle_0 &=& 4 N_c\int
\frac{d^4p}{(2\pi)^4}
\left(\frac{m_f+M_f(p)}{(p^2+(m_f+M_f(p))^2)^2}-
\frac{m_f}{(p^2+m_{f}^2)^2}\right)
\nonumber\\
&-&4 N_c\int \frac{d^4p}{(2\pi)^4} \frac{M_f
F(p)F'(p)p}{(p^2+(m_f+M_f(p))^2)^2},
\label{chi-final-integ}
\eea
where the quark condensate in the chiral limit plays a role of
the normalization and is given by:
\begin{equation}
\langle i\psi^\dagger \psi \rangle_0 = 4 N_c\int
\frac{d^4p}{(2\pi)^4} \frac{M_0 F^2 (p)}{p^2 + M_0^2 F^4 (p)
}\approx 0.017 \,{\rm GeV}^3,
\end{equation}
The infrared region of the first integral in Eq.\re{chi-final-integ}
brings out the contribution of order ${\cal O}(m_f\,\ln\,m_f)$ which
is almost model-independent.  In order to calculate the magnetic 
susceptibility explicitly, we first consider two different regions
while integrating: $0<p<1 \,{\rm GeV}$ and 
$1\,{\rm GeV}< p < \infty$.  Then we are able to evaluate a part of
the free quark analytically. The second integral in
Eq.~\re{chi-final-integ} is related to the nonlocal contribution
without which the vector current is not conserved, i.e. the
Ward-Takahashi identity is broken.  It arises from the nonlocal
quark-quark interaction and is called {\em nonlocal current}. Summing
all contributions, we obtain the final result for the magnetic
susceptibility of the QCD vacuum at $q^2=0$:  
\begin{equation}
\chi_f = \chi_f^{\rm local} + \chi_f^{\rm nonlocal}
\label{chi-final}
\end{equation}
with
\begin{eqnarray}
\chi_f^{\rm local} \langle i\psi_f^\dagger \psi_f \rangle_0 &=&
\left(29+0.103 m_f + \frac{3\, m_f}{2\pi^2} \ln\,\frac{m_f}{1000\,
{\rm MeV}}\right)
\,[{\rm MeV}], \label{eq:final1} \\
\chi_f^{\rm nonlocal} \langle i\psi_f^\dagger \psi_f \rangle_0 &=&
\left(18 - 0.095 m_f \right)
\,[{\rm MeV}].
\label{eq:final2}
\end{eqnarray}
In principle, we can also derive the magnetic susceptibility at 
finite $q^2$, i.e. the form factor of the magnetic susceptibility of
the QCD vacuum.  


\begin{figure}[t]
  \centering
\includegraphics[height=7cm]{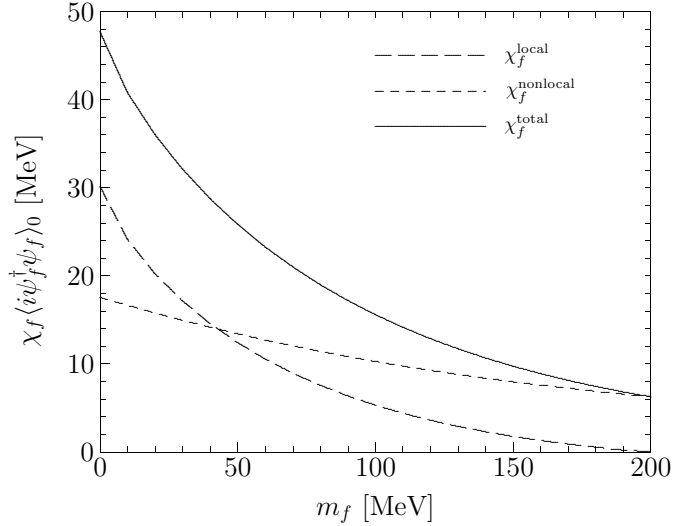}
  \caption{The $m_f$ dependence of the magnetic susceptibility with
    the form factor of the zero-mode type (Eq. (\ref{eq:ff}))
    used.  The long-dashed curve draws the contribution of the local
    vector current, while the short-dashed one depicts that of the
    nonlocal one.  The solid curve represents the total magnetic
    susceptibility as a function of the current quark mass.}
  \label{fig:1}
\end{figure}

{\bf 5.}  Figure~\ref{fig:1} shows the magnetic susceptibility
$\chi_f\langle i\psi_f^\dagger \psi_f\rangle_0$ as a function of
the current quark mass, with the zero-mode form factor
in Eq.~(\ref{eq:ff}) employed.  It is a monotonically decreasing
function as $m_f$ increases, since the 
third term with $m_f {\rm ln}\, m_f $ in Eq.~(\ref{eq:final1})
suppresses the first and second ones noticeably as $m_f$ gets
larger.  In fact, its value is around $-4.0\,{\rm MeV}$ for up
and down quark masses $m_{u,d}\simeq 5\,{\rm MeV}$, whereas for
the strange quark mass $m_s\simeq 200$ MeV it turns out to be
$-49\,{\rm MeV}$.  Thus, it almost cancels the leading term in
the case of the strange magnetic susceptibility $\chi_s$.
Moreover, the nonlocal part is responsible for about $40\%$ of
the total magnetic susceptibility in the chiral limit and
diminishes linearly but slowly as $m_f$ increases.  It implies
that it is essential to treat the vector current properly in the
presence of the nonlocal interaction, in particular, in the case
of the magnetic susceptibility.  As a result, the magnetic
susceptibility for the up and down quarks is: $\chi_{u,d}=40\sim
45$ MeV, whereas for the strange quark we obtain $\chi_{s}=6\sim 10$ 
MeV.  As for the up and down quarks, the present result is comparable
to that of Ref.~\cite{Belyaev:ic,Balitsky:aq,Ball:2002ps}. 

\begin{figure}[t]
  \centering
\includegraphics[height=7cm]{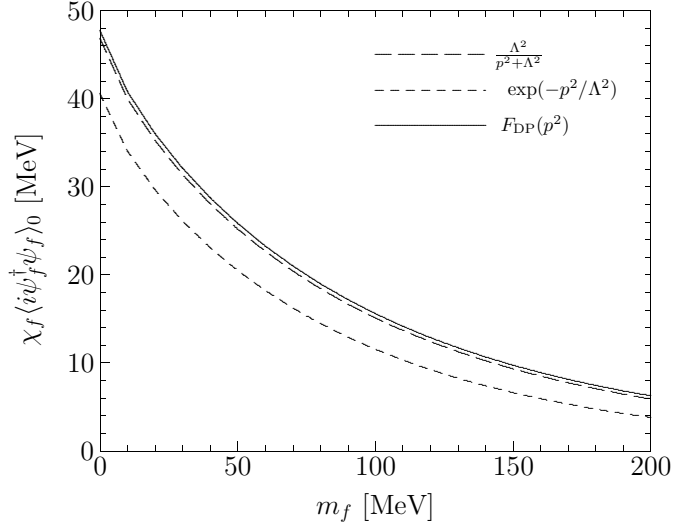}
\caption{A comparison between different types of the form
factors.  The solid curve is drawn with the zero-mode
form factor (\ref{eq:ff}), while the long-dashed one depicts the
dipole-type form factor. The short-dashed one is for the Gaussian
form factor.}
  \label{fig:2}
\end{figure}

A comparison between different types of the form factors is
depicted in Fig.~\ref{fig:2}.  As already discussed in
Refs.~\cite{Franz:1999ik,Choi:2003cz,Lee:2004tr}, the exponential
type of the form factor yields the smallest result as usual.  It is
interesting to see that the dipole type presents a very similar
result for the magnetic susceptibility to that with the 
zero-mode form factor (\ref{eq:ff}).  It is also found that
the dependence of the magnetic susceptibility on a type of the
form factors is not sensitive.


{\bf 6.} In the present work, we have investigated the magnetic
susceptibility of the QCD vacuum within the framework of the
nonlocal chiral quark model from the instanton vacuum.  We first
have constructed the effective partition function in the presence
of the external vector and tensor fields as well as the current
quark mass.  It was shown that for the constant electromagnetic
field strength the present result is independent of the path which is
selected for the gauge connection.  We found that the result is a
smoothly decreasing function of $m_f$: For the up and down quarks 
$m_{u,d}\simeq 5\,{\rm MeV}$ we have $\chi_{u,d}\langle i
\psi_{u,d}^\dagger \psi_{u,d}\rangle_0= 40\sim 45 \,{\rm MeV} $
and for the strange quark we obtain $\chi_{s}\langle i
\psi_{s}^\dagger \psi_{s}\rangle_0= 6\sim 10 \,{\rm MeV}$.  As
for the up and down quarks the present results are comparable to
the estimate of the vector dominance and QCD sum rule
approach~\cite{Belyaev:ic,Balitsky:aq,Ball:2002ps}.


The present work is supported by Korea Research Foundation Grant
(KRF-2003-041-C20067).  MM acknowledges the support of the Brain
Pool program 2004 which makes it possible for him to 
visit Pusan National University.

\end{document}